\documentclass[twocolumn,prb,noshowpacs]{revtex4}
\usepackage{epsfig}
\usepackage{caption}
\usepackage{lscape}
\usepackage{bm}
\usepackage{amsfonts}
\usepackage{amsmath}
\usepackage{amssymb}
\usepackage{graphicx}

\input{tcilatex}

\begin{document}

\title{Spin transverse force and quantum transverse transport}
\author{Bin Zhou and Shun-Qing Shen}
\address{\textit{Department of Physics, and Center for Theoretical and
Computational Physics, The University of Hong Kong, Pokfulam Road, Hong
Kong, China}}

\begin{abstract}
We present a brief review on spin transverse force, which exerts on the spin
as the electron is moving in an electric field. This force, analogue to the
Lorentz force on electron charge, is perpendicular to the electric field and
spin current carried by the electron. The force stems from the spin-orbit
coupling of electrons as a relativistic quantum effect, and could be used to
understand the \textit{Zitterbewegung} of electron wave packet and the
quantum transverse transport of electron in a heuristic way.
\end{abstract}

\maketitle

\section{Introduction}

In electrodynamics it is well known that a magnetic field would exert a
transverse force, i.e. the Lorentz force, on an charged particle if it is
moving. This Lorentz force can lead to a lot of fundamental phenomena such
as the Hall effect in solid.\cite{Hall} The interaction of the electron spin
in the electromagnetic field behaves as if the spin is a gauge charge and
the interaction is due to the SU(2) gauge field.\cite{Anandan89} Essentially
the electron spin is an intrinsic quantum variable, not a just classical
tiny magnetic moment. If we want to manipulate and control quantum spin
states, a natural question raises: what type of force exerts on a moving
spin in an electric field? In the recent work\cite{Shen05prl,Li05prb}, it
was found that an electric field exerts a transverse force on an electron
spin if it is moving and the spin is projected along the electric field. The
force stems from the spin-orbit coupling which can be derived from the Dirac
equation for an electron in a potential in the non-relativistic limit or the
Kane model with the $\mathbf{k}\cdot \mathbf{p}$ coupling between the
conduction band and valence band. It is heuristic to understand that the 
\textit{Zitterbewegung} of electronic wave packet is driven by the spin
transverse force on a moving spin. Based on the concept of spin force
balance, a relation between spin current and spin polarization is given. The
role of spin transverse force is also discussed in the anomalous Hall
effect, the spin Hall effect and its reciprocal effect driven by the pure
spin current in semiconductors. It was shown that the Kubo formula can be
formalized in terms of the spin force. Some authors also discussed the spin
force in the systems with Rashba coupling, and its relation with the spin
transport by means of numerical simulation,\cite{Li05prb,Nikolic05prb} the
Boltzman equations,\cite{Adagideli05} and the continuity equation of the
momentum current.\cite{Wang06prl} The forces induced by the Yang-Mills filed
such as the Rashba and Dresselhaus fields, the sheer strain field acting on
spin and spin current were also derived.\cite{Jin06jpa}

\section{Spin Transverse force}

We consider an electron in a confining potential $V$ and a vector potential $%
\mathbf{A}$ for a magnetic field, $\mathbf{B}=\nabla \times \mathbf{A}$. In
the relativistic quantum mechanics, the Dirac equation determines the
behaviors of electron. In the nonrelativistic limit, neglecting the higher
order terms of expansion, the Dirac equation is reduced to the Schr\"{o}%
dinger equation by introducing the spin-orbit coupling and the Zeeman
exchange coupling,%
\begin{equation}
H\approx \frac{\left( \mathbf{p}+\frac{e}{c}\mathbf{A}\right) ^{2}}{2m}%
+V_{eff}+\mu _{B}\sigma \cdot B+\frac{\hbar \left( \mathbf{p}+\frac{e}{c}%
\mathbf{A}\right) }{4m^{2}c^{2}}\cdot \left( \sigma \times \nabla V\right) 
\text{,}  \label{soc}
\end{equation}%
where $m$ and $e$ are the electron mass and charge, respectively, $c$ is the
speed of light, $\sigma $ are the Pauli matrices, $\mu _{B}=e\hbar /2mc$ and 
$V_{eff}=V+\frac{\hbar ^{2}}{8m^{2}c^{2}}\nabla ^{2}V$. From the Heisenberg
motion equation, we can obtain the quantum mechanical analogue of Newton's
second law, known as the Erhrenfest theorem, 
\begin{equation}
F=m\frac{d\mathbf{v}}{dt}=\frac{m}{i\hbar }\left[ \mathbf{v},H\right] =-%
\frac{m}{\hbar ^{2}}\left[ \left[ \mathbf{r},H\right] ,H\right] .
\label{force}
\end{equation}%
Of course there is no concept of force in quantum mechanics, this is just an
operator equation. The physical meaning of force is contained in the
expectation value in a quantum state. The quantum mechanical version of the
force $F$ comprises three parts, \textit{i.e.}, Lorentz force $F_{h}$, spin
electromagnetic force $F_{g}$ and spin transverse force $F_{f}$. The form of
Lorentz force $F_{h},$ which has its counterpart in the classic limit of $%
\hbar \rightarrow 0$, is 
\begin{equation}
F_{h}=-\frac{e}{c}\left( \mathbf{v}\times \mathcal{B-B}\times \mathbf{v}%
\right) /2-\nabla \left( V_{eff}+\mu _{B}\sigma \cdot B\right) ,
\end{equation}%
where $\mathcal{B}=\mathbf{B}+\nabla \times \mathcal{A}$, with $\mathcal{A}=%
\frac{\hbar }{4mce}\sigma \times \nabla V$. It indicates clearly that $%
\mathcal{A}$ plays a role of a SU(2) gauge vector potential. Recently it is
proposed that the force can generate a pure spin current if we assume $%
\nabla B_{z}$ is a constant.\cite{Shi06prl} The spin electromagnetic force $%
F_{g}$ is written as%
\begin{equation}
F_{g}=\frac{\mu _{B}}{2mc^{2}}[\sigma (\mathbf{B}\cdot \nabla V)-\mathbf{B}%
(\sigma \cdot \nabla V)].
\end{equation}%
It is non-zero only when the electric and magnetic fields coexist, as
suggested by Anandan and others.\cite{Anandan89} The spin transverse force $%
F_{f}$ is derived as 
\begin{equation}
F_{f}=\frac{\hbar }{8m^{2}c^{4}}(\sigma \cdot \nabla V)(\mathbf{v}\times
\nabla V),
\end{equation}%
which comes from the SU(2) gauge potential or spin-orbital coupling. As the
force is related to the Planck constant it has no counterpart in classic
mechanics and is purely quantum mechanic effect. The spin transverse force
for a single electron on a quantum state can be written in a compact form,%
\begin{equation}
\left\langle F_{f}\right\rangle =\frac{e^{2}\left\vert \mathcal{E}%
\right\vert }{4m^{2}c^{4}}\mathbf{j}_{s}^{\mathcal{E}}\times \mathcal{E},
\label{spinforce}
\end{equation}%
where the spin current is defined conventionally, $\mathbf{j}_{s}^{\mathcal{E%
}}=\frac{\hbar }{4}\left\langle \{\mathbf{v},\sigma \cdot \mathcal{E}%
/\left\vert \mathcal{E}\right\vert \}\right\rangle $. The force is
proportional to the square of electric field $\mathcal{E}$ and the spin
current which polarization is projected along the field. It is important to
note that an electron in a spin state perpendicular to the electric field
will not experience any force. Comparing with a charged particle in a
magnetic field, $\mathbf{j}_{c}\times \mathbf{B}$, where $\mathbf{j}_{c}$ is
a charge current density, the spin force is nonlinear to the electric field
and depends on the spin state of electron.

\section{Applications of spin transverse force}

\textit{Zitterbewegung of wave package-}The rapid oscillation of the
electron wave packet is known in literatures as the \textit{Zitterbewegung}
of electron as a relativistic quantum mechanical effect, which is physically
regarded as a result of admixture of the positron state in electron wave
packet.\cite{Schrodinger30} However, this effect is usually expected to be
measured in high energy physics, but fails to do experimentally. Note that
the effective models describing the band structure of III-V semiconductors
are similar to the Dirac equation such that some authors proposed that the
experimental observation of \textit{Zitterbewegung }may be more realistic in
semiconductors with spin-orbit coupling rather than in high energy physics.
Recently Schliemann \textit{et al} proposed that \textit{Zitterbewegung} can
be observed in both $n$- and $p$-doped III-V zinc-blende semiconductor
quantum wells.\cite{Schliemann05prl} In the $p$-doped three-dimensional bulk
semiconductors described by the Luttinger model will generate the \textit{%
Zitterbewegung} as calculated by Jiang \textit{et al}.\cite{Jiang05prb} The 
\textit{Zitterbewegung }of\textit{\ }electrons in three-dimensional bulk
III-V semiconductors was also discussed by Zawadzki.\cite{Zawadzki05prb}
Katsnelson analyzed the oscillatory motion of the electron related to the 
\textit{Zitterbewegung }in graphene which is a gapless semiconductor with
massless Dirac energy spectrum.\cite{Katsnelson06EPJB} Very recently, a
unified treatment of \textit{Zitterbewegung }for spintronic, graphene, and
superconducting systems was presented.\cite{Cserti06xxx} Readers are
referred to see other recent worked devoted to \textit{Zitterbewegung }%
phenomena in references.\cite{Nikolic05prb,Lee05prb}

In the recent work,\cite{Shen05prl} a heuristic picture is given to
understand that the \textit{Zitterbewegung} of electronic wave packet is
driven by the spin transverse force on a moving spin. Here we consider the
motion of an electron confining in a two-dimensional plane subjected to a
perpendicular electric field, without loss of generality, and assume that
the spin-orbit coupling provides an effective magnetic field along the $y$
direction. In this case, only the spin transverse force $F_{f}\ $exerts on
the spin while the Lorentz force $F_{h}$ and spin electromagnetic force $%
F_{g}$ vanish. Because of the spin-orbit coupling the the electron spin
precesses in the spin $x$-$z$ plane.\cite{Shen04prb} The spin $\sigma
_{z}(t) $ varies with time and the spin current is always along the $x$
direction. As a result the spin transverse force is always perpendicular to
the $x$ direction. If the initial state is polarized along $y$ direction the
electron spin does not vary with time as it is an energy eigenstate of the
system. In this case the spin current carried by the electron is always
zero, as a result, the spin transverse force is also zero. If the initial
state is along the spin $z$ direction, the spin state will evolve with time
as the the state is not the energy eigenstate. It can be understood that the
spin precession makes the spin current whose polarization is projected along
the electric field changes with time such that the spin force along the $y$
direction also oscillates. This force will generate a non-zero velocity of
electron oscillating along the $y$ direction, and then the trajectory
oscillates with the time.

Though the spin transverse force on a moving spin is very analogous to the
Lorentz force on a moving charge, because of spin precession, its effect is
completely different with the motion of a charged particle in a magnetic
field, where the amplitude of the Lorentz force is constant and the charged
particle will move in a circle. The \textit{Zitterbewegung} of the
electronic wave package near the boundary will cause some edge effect as
shown in recent numerical calculations.\cite{edge} The edge effect is
determined by the electron momentum. The smaller the momentum, the larger
the edge effect.

\textit{Spin force balance}-Intrinsic spin Hall effect has been received a
great deal of attention recently since the works by Murakami \textit{et al}%
\cite{Murakami03sci} and Sinova \textit{et al}\cite{Sinova04prl} that a spin
Hall current can be produced by an electric field in $p$-doped III-V
semiconductors and the two-dimensional electron gas with Rashba coupling.
Vertex correction and numerical calculations show that disorder can cancel
spin Hall effect in the two-dimensional electron gas with Rashba coupling.%
\cite{Inoue04prb} On the other hand, if the exchange coupling is taken into
account, spin Hall effect should survive.\cite{Shen04prl} In fact, the
result can be derived based on the concept of spin force balance.\cite%
{Zhou06prb} We consider an effective Hamiltonian for a two-dimensional
ferromagnetic system with Rashba coupling, $H=p^{2}/\left( 2m^{\ast }\right)
+\lambda (p_{x}\sigma _{y}-p_{y}\sigma _{x})+h_{0}\sigma _{z}$, where $%
m^{\ast }$ is the effective mass of conduction electron and the exchange
field $h_{0}$ due to the magnetic impurities. The spin-orbit coupling
induces the spin transverse force $F_{f}=\frac{4m^{\ast 2}\lambda ^{2}}{%
\hbar ^{2}}J_{s}^{z}\times \hat{z}$, and the spin electromagnetic force $%
F_{g}=-\frac{2m^{\ast }\lambda h_{0}}{\hbar }\left[ \sigma _{x}\hat{x}%
+\sigma _{y}\hat{y}\right] $. If the disorder potential $V_{disorder}$ is
taken into account, in a steady state, the spin force must reach at balance, 
\begin{equation}
\frac{1}{i\hbar }\left\langle \left[ \frac{e}{c}\mathcal{A},H+V_{disorder}%
\right] \right\rangle =\left\langle F_{f}+F_{g}\right\rangle =0.
\end{equation}%
This result is independent of the non-magnetic disorder and interaction
because the spin gauge field commutes with non-magnetic potential $%
V_{disorder}$. From the spin force balance we have a relation between spin
current and spin polarization, $\left\langle J_{x}^{z}\right\rangle =+\frac{%
\hbar h_{0}}{2m^{\ast }\lambda }\left\langle \sigma _{y}\right\rangle $, and 
$\left\langle J_{y}^{z}\right\rangle =-\frac{\hbar h_{0}}{2m^{\ast }\lambda }%
\left\langle \sigma _{x}\right\rangle $. It is obvious that the spin Hall
current vanishes in the case of $h_{0}=0$, while if $h_{0}\neq 0$ spin Hall
current should survive.

\textit{Anomalous Hall effect}-The spin transverse force can be also
regarded a driven force of an anomalous Hall effect\cite{AHE}. Here we give
a clear picture for anomalous Hall effect in ferromagnetic metal. When an
external electric field is applied along the $x$ axis, it will circulate an
electric current $J_{c,x}$, and also a spin current $J_{x}^{z}$ since the
charge carriers are partially polarized. The spin-orbit coupling exerts a
spin transverse force on the spin current, $J_{x}^{z}$, and generate a drift
velocity or the anomalous Hall current $J_{c,y}$. From the Rashba coupling
the spin polarization tends to be normal to the momentum or electric
current. The electric current $J_{c,x}$ along $E$ induces a non-zero $%
\left\langle \sigma _{y}\right\rangle $ and the anomalous Hall current $%
J_{c,y}$ induces non-zero $\left\langle \sigma _{x}\right\rangle $. These
non-zero spin polarization maintains the balance of spin transverse force,
and further a non-zero spin current in a steady state. Thus the anomalous
electronic transverse transport is robust against the disorder in the
ferromagnetic metals and semiconductors.

\textit{Spin resolved charge Hall effect}-The spin transverse force can be
applied understand the spin-resolved Hall effect.\cite{Bulgakov99prl,Li05prb}
When a spin current is injected into 2DEG with the spin-orbit coupling, the
spin-orbit coupling exerts the spin transverse force on the spin current,
and drives electrons to form a charge Hall current perpendicular to the spin
current. The\ injected spin current which can be generated in various ways,
such as the ac-magnetic field,\cite{Shangguan} the spin force\ $\nabla (%
\mathbf{\mu }\cdot \mathbf{B),}$\cite{Shi06prl} and circularly or linearly
polarized light injection.\cite{Sipe,Hankiewicz05prb,Li06apl}

\textit{Spin transverse force and Kubo formula}-The quantum transverse
transport of electrons caused by a weak electric field $\mathbf{E}$ can be
calculated within the framework of linear response theory. Based on the
concept of spin force, the Kubo formula of linear response theory can be
re-formalized.\cite{Zhou06prb} We consider an effective Hamiltonian for
electrons with spin 1/2, $H=p^{2}/\left( 2m^{\ast }\right) +\sum_{\alpha
=x,y,z}d_{\alpha }(p)\sigma _{\alpha }$, where $d_{\alpha }(p)$ are the the
momentum-dependent coefficients which describes the spin-orbit interactions
and exchange interaction. The spin force is derived as $F_{j}=\frac{2m^{\ast
}}{\hbar }\mathbf{\sigma }\cdot (\frac{\partial \mathbf{d}}{\partial p_{j}}%
\times \mathbf{d})$. In general, for any observable $O$, its linear response
to an external electric field $\mathbf{E}$ has the formula, $\left\langle
O_{i}\right\rangle =\chi _{ij}E,$ and 
\begin{equation}
\chi _{ij}=-\frac{e\hslash ^{2}}{16m^{\ast }\Omega }\sum_{p}\frac{\left(
f_{p,-}-f_{p,+}\right) }{d^{3}}\mathrm{{Tr}\left[ \{F_{j},O_{i}\}\right] ,}
\end{equation}%
where $d=\sqrt{d_{x}^{2}+d_{y}^{2}+d_{z}^{2}}$ and $f_{p,\pm }$ are the
Dirac-Fermi distribution functions.

\section{Conclusion}

In conclusion, compared with the Lorentz force\ brought by the magnetic
field upon a charged particle, the spin-orbit coupling produces a spin
transverse force on a moving electron spin. It has no classical counterpart
as the coefficient is divided by $\hbar $, but it reflects the tendency of
spin asymmetric scattering of a moving electron subject to the spin-orbit
coupling. It stems from the spin-orbit coupling as a relativistic quantum
mechanical effect. However, this quantity may provide an new way to
understand the spin transport in semiconductors

This work was supported by the Research Grant Council of Hong Kong under
Grant No.: HKU 7042/06P.


\begin{thebibliography}{99}
\bibitem{Hall} The Hall Effect and its Applications, edited by C. L. Chien
and C. R. Westgate (Plenum, New York, 1980).

\bibitem{Anandan89} J. Anandan, \textit{Phys. Lett. A} \textbf{138}, 347
(1989); J. Anandan, \textit{Phys. Lett. A} \textbf{152}, 504 (1991); R. C.
Casella and S. A. Werner, \textit{Phys. Rev. Lett.} \textbf{69}, 1625; A. V.
Balatsky and B. L. Altshuler, \textit{Phys. Rev. Lett}. \textbf{70}, 1678
(1993); C. M. Ryu, \textit{Phys. Rev. Lett}. \textbf{76}, 968 (1996); T.
Choi, \textit{Mod. Phys. Lett. B} \textbf{20}, 189 (2006).

\bibitem{Shen05prl} S. Q. Shen, \textit{Phys. Rev. Lett.} \textbf{95},
187203 (2005).

\bibitem{Li05prb} J. Li, L. Hu, and S. Q. Shen, \textit{Phys. Rev. B} 
\textbf{71}, 241305 (2005).

\bibitem{Nikolic05prb} B. K. Nikoli\'{c}, L. P. Z\^{a}rbo, and S. Welack, 
\textit{Phys. Rev. B} \textbf{72}, 075335 (2005).

\bibitem{Adagideli05} \.{I}. Adagideli and G. E. W. Bauer, \textit{Phys.
Rev. Lett.} \textbf{95}, 256602 (2005).

\bibitem{Wang06prl} Y. Wang, K. Xia, Z. B. Su, and Z. Ma, \textit{Phys. Rev.
Lett.} \textbf{96}, 066601 (2006).

\bibitem{Jin06jpa} P. Q. Jin, Y. Q. Li, and F. C. Zhang, \textit{J. Phys. A:
Math. Gen.} \textbf{39}, 7115 (2006).

\bibitem{Shi06prl} J. Shi, P. Zhang, D. Xiao, and Q. Niu, \textit{Phys. Rev.
Lett.} \textbf{96}, 076604 (2006).

\bibitem{Schrodinger30} E. Schrodinger, \textit{Sitzungsber. Preuss. Akad.
Phys. Math. Kl.} \textbf{24}, 418 (1930); H. Feshbach and F. Villars, 
\textit{Rev. Mod. Phys.} \textbf{30}, 24 (1958).

\bibitem{Schliemann05prl} J. Schliemann, D. Loss, and R. M. Westervelt, 
\textit{Phys. Rev. Lett.} \textbf{94}, 206801 (2005); \textit{Phys. Rev. B} 
\textbf{73}, 085323 (2006).

\bibitem{Jiang05prb} Z. F. Jiang, R. D. Li, S. C. Zhang, and W. M. Liu, 
\textit{Phys. Rev. B} \textbf{72}, 045201 (2005).

\bibitem{Zawadzki05prb} W. Zawadzki, \textit{Phys. Rev. B} \textbf{72},
085217 (2005).

\bibitem{Katsnelson06EPJB} M. I. Katsnelson, \textit{Eur. Phys. J. B} 
\textbf{51}, 157 (2006).

\bibitem{Cserti06xxx} J. Cserti and G. D\'{a}vid, cond-mat/0604526.

\bibitem{Lee05prb} M. Lee and C. Bruder, \textit{Phys. Rev. B} \textbf{72},
045353 (2005); P. Brusheim and H. Q. Xu, cond-mat/0512502; W. Zawadzki,
cond-mat/0510184; T. M. Rusin and W. Zawadzki, cond-mat/0605384; J. Tworzyd%
\l o, B. Trauzettel, M. Titov, A. Rycerz, and C. W. J. Beenakker, \textit{%
Phys. Rev. Lett.} \textbf{96}, 246802 (2006); B. Trauzettel, Ya. M. Blanter,
and A. F. Morpurgo, cond-mat/0606505.

\bibitem{Shen04prb} S. Q. Shen, \textit{Phys. Rev. B} \textbf{70}, 081311
(2004).

\bibitem{edge} A. Reynoso, G. Usaj, M. J. S\'{a}nchez, and C. A. Balseiro, 
\textit{Phys. Rev. B} \textbf{70}, 235344 (2004).

\bibitem{Murakami03sci} S. Murakami, N. Nagaosa, and S. C. Zhang, \textit{%
Science} \textbf{301}, 1348 (2003).

\bibitem{Sinova04prl} J. Sinova, D. Culcer, Q. Niu, N. A. Sinitsyn,
T.Jungwirth, and A. H. MacDonald, \textit{Phys. Rev. Lett.} \textbf{92},
126603 (2004).

\bibitem{Inoue04prb} J. I. Inoue, G. E. W. Bauer and L. W. Molenkamp, 
\textit{Phys. Rev. B} \textbf{70}, 041303(R) (2004); E. G. Mishchenko, A. V.
Shytov, and B. I. Halperin, \textit{Phys. Rev. Lett.} \textbf{93}, 226602
(2004); K. Nomura, J. Sinova, T. Jungwirth, Q. Niu, and A. H. MacDonald, 
\textit{Phys. Rev. B} \textbf{71}, 041304 (2005); D. N. Sheng, L. Sheng, Z.
Y. Weng, and F. D. M. Haldane, \textit{Phys. Rev. B} \textbf{72}, 153307
(2005).

\bibitem{Shen04prl} S. Q. Shen, M. Ma, X. C Xie, and F. C. Zhang, \textit{%
Phys. Rev. Lett.} \textbf{92}, 256603 (2004); S. Q. Shen, Y. J. Bao, M. Ma,
X. C Xie, and F. C. Zhang, \textit{Phys. Rev. B} \textbf{71}, 155316 (2005);
Y. J. Bao, H. B. Zhuang, S. Q. Shen, and F. C. Zhang, \textit{Phys. Rev. B} 
\textbf{72}, 245323 (2005).

\bibitem{Zhou06prb} B. Zhou, L. Ren, and S. Q. Shen, \textit{Phys. Rev. B} 
\textbf{73}, 165303 (2006).

\bibitem{AHE} R. Karplus and J. M. Luttinger, \textit{Phys. Rev.} \textbf{95}%
, 1154 (1954); J. Smit, \textit{Physica} (Amsterdam) \textbf{21}, 887
(1955); L. Berger, \textit{Phys. Rev. B} \textbf{2}, 4559 (1970).

\bibitem{Bulgakov99prl} E. N. Bulgakov, K. N. Pichugin, A. F. Sadreev, P. St%
\v{r}eda, and P. \v{S}eda, \textit{Phys. Rev. Lett}. \textbf{83}, 376 (1999).

\bibitem{Shangguan} M. Shangguan, Q. F. Sun, J. Wang, and H. Guo, \textit{%
Phys. Rev. B} \textbf{73}, 125349 (2006).

\bibitem{Sipe} E. Ya. Sherman, A. Najmaie, and J. Sipe, \textit{Appl. Phys.
Lett.} \textbf{86}, 122103 (2005).

\bibitem{Hankiewicz05prb} E. M. Hankiewicz, J. Li, T. Jungwirth, Q. Niu, S.
Q. Shen, and J. Sinova, \textit{Phys. Rev. B} \textbf{72}, 155305 (2005).

\bibitem{Li06apl} J. Li, X. Dai, S. Q. Shen, and F. C. Zhang, \textit{Appl.
Phys. Lett.} \textbf{88}, 162105 (2006).
\end{thebibliography}
\end{document}